\documentclass[10pt,a4paper]{article}

\usepackage{natbib}
\usepackage[dvips]{graphicx}
\usepackage{epsfig}  
\usepackage{amsmath}
\usepackage{subfigure}

\author{Andri Mirzal\\Graduate School of Information Science and Technology,\\ Hokkaido University, Kita 14 Nishi 9, Kita-Ku,\\
Sapporo 060-0814, Japan\\}

\title{On the Relationship between Trading Network and WWW Network: A Preferential Attachment Perspective}

\begin{document}

\maketitle

\begin{abstract}
This paper describes the relationship between trading network and WWW network from preferential attachment mechanism perspective. This mechanism is known to be the underlying principle in the network evolution and has been incorporated to formulate two famous web pages ranking algorithms, PageRank and HITS. We point out the differences between trading network and WWW network in this mechanism, derive the formulation of HITS-based ranking algorithm for trading network as a direct consequence of the differences, and apply the same framework when deriving the formulation back to the HITS formulation that turns to become a technique to accelerate its convergences.
\end{abstract}

\section{Introduction} \label{sec1}
\noindent The researches on the analysis of preferential attachment and network structure can be dated back to 50's with the work of \cite{1}, where the authors presented the first systematic study of a class of networks known as \emph{random graphs}. Actually the study of graph itself has a long history in mathematics as Euler introduced the using of vertices and edges to model the famous K\"onigsberg bridge problem in 1736. However, different from the classical studies, the modern network studies have some interesting additional features: (1) focusing on much larger problems that can contain million vertices so it is natural to consider statistical properties of the networks, (2) dealing with real networks like Internet topology (\citealp{4}), WWW network (\citealp{5,6}), metabolic networks (\citealp{7,8}), scientific collaboration networks (\citealp{26,9,14,15}), and epidemic spreading networks (\citealp{10,11,12,13}) among others, and (3) studying dynamical properties of the networks as many real networks are not static entities, but grow according to some rules (\citealp{16,14,15}).

The foundation of modern random graph theory which focuses on structure and statistical properties of very large random graphs was set by Erd\H os and Reny\'i (\citealp{17,18,19}). The random graph is a very influential model in modeling the real networks because it can describe many phenomena including phase transition, short paths between most of vertex pairs, and the existence of a giant component. Prior to the finding of scale-free network (\citealp{25}), many network designs were based on the random graph model, including Internet data protocols design (\citealp{20}) and social network experiments setting (\citealp{22,21}).

The first widely known challenge to the random graph came from the study of WWW network topology by Albert, Jeong, and Barab\'asi\protect\footnote{We must note here actually Price (\citealp{26}) is the first person to show that real networks, scientific collaboration networks, are following power-law degree distribution, and his finding was long before the works of Barab\'asi et al.~(\citealp{14,15}). Interestingly, Price didn't seem to know about the famous random graph model of Erd\H os and Reny\'i, and also Barab\'asi et al.~didn't know about the previous work of Price when studying scientific collaboration networks.}  (\citealp{5,25}). While the first paper's main focus is in the short path between any pair of pages that supports the finding of \cite{23}, the second paper is the first to explicitly challenge the effectiveness of the random graph in modeling the real networks by showing experimentally the ubiquitous existence of power-law degree distributions ($p_{k} \propto k^{-\gamma}$, where $p_{k}$ is the probability of vertices with degree $k$, and $\gamma$ is the exponent that usually in the range 2-3 in the real networks) in a variety of real networks including WWW network, scientific collaboration networks, and film actors networks that led to the famous scale-free hypothesis.

The power-law degree distributions found in many real networks are considered to be the most important remark that shows the discrepancy between random graph prediction of the degree distribution (which should follow Poisson distribution, $p_{k} \propto m^{k}e^{-m}/k!$, where $m$ is the mean degree of vertices) and the real situation. This discrepancy outclasses other good predictions made by the random graph including short path, diameter of the graph, phase transition, and size of the giant component, and generates a very large number of scientific publications on such networks vary from mathematics, physics, computer science, economics to sociology. And in turn creates a new field of study: \emph{complex networks}. 
\begin{figure}[t]
 \begin{center}
 
  \subfigure[Poisson distribution]{
   \includegraphics[width=0.45\textwidth]{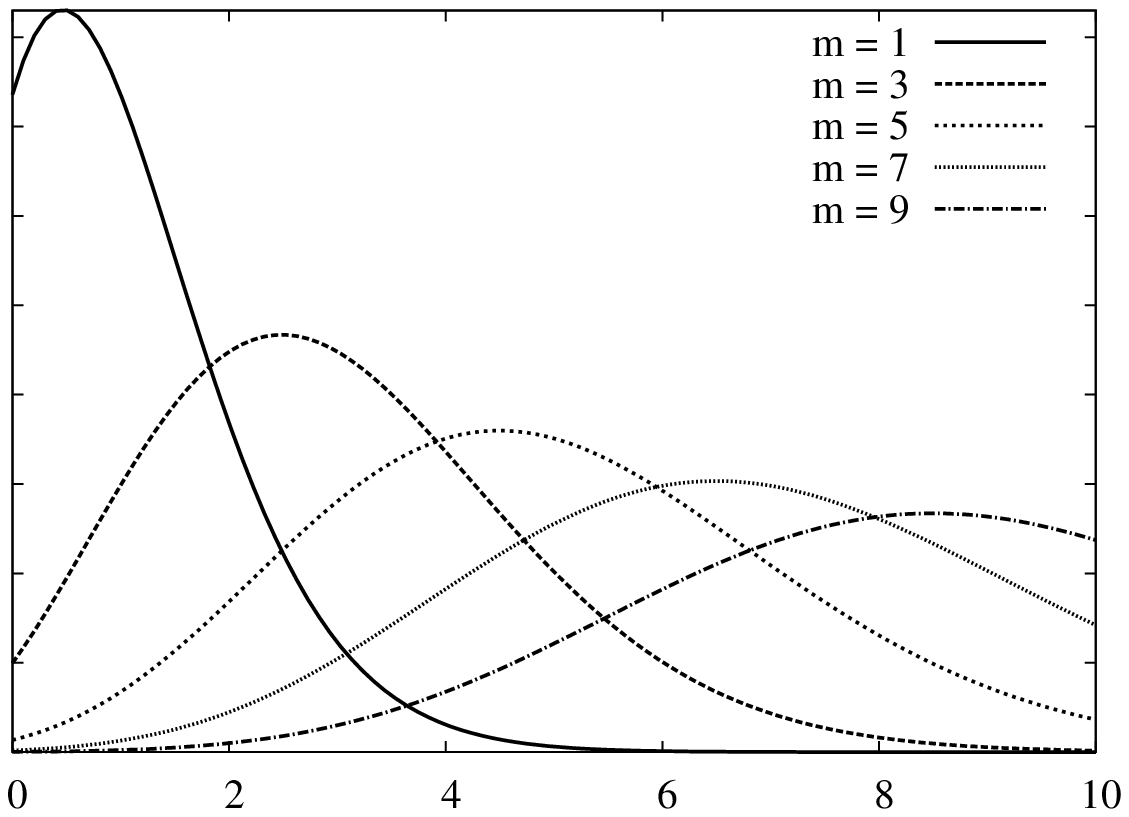}
   \label{poisson}
  }
  \subfigure[Power-law distribution]{
   \includegraphics[width=0.45\textwidth]{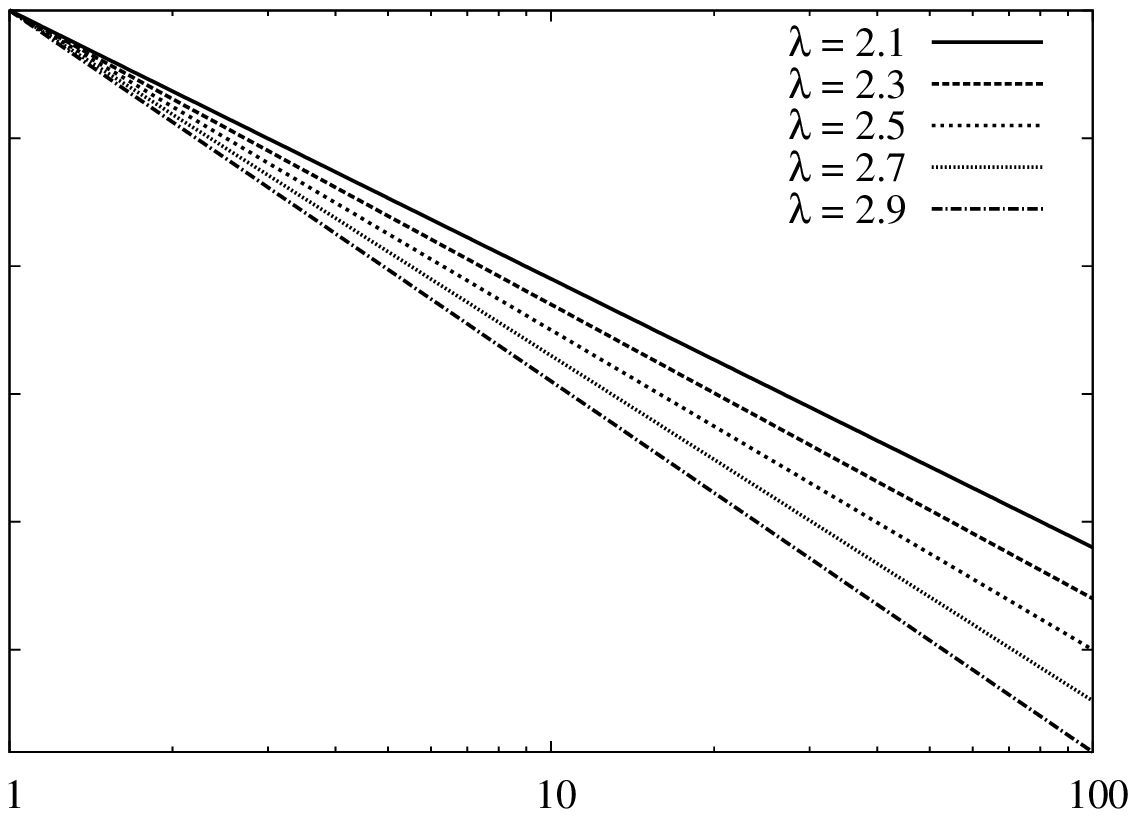}
   \label{powerlaw}
  }
  \caption{Poisson and power-law distribution plots for several $m$ and $\lambda$ respectively. Note that x-axis is the degree $k$, y-axis is the probability $p_k$, and (b) is in log-log scale.}
  \label{fig1}
 \end{center}
\end{figure}

There are several fundamental reasons behind the curiosity in the ubiquity of the scale-free phenomenon. We enlist some of them here:
\begin{enumerate}
\item Different from the Poisson distribution, mean value and standard deviation of the power-law distribution doesn't imply centrality and data dispersion. As we know, these metrics can be very useful in describing a distribution without having to plot it. But in the case of the power-law, these metrices can be misleading because the mean value doesn't reflect the centrality and the standard deviation doesn't tell us about the range where most of the data lies.
\item There is no peak value in the power-law, $p_k$ decreases monotonically as $k$ increases.
\item The power-law has a large tail that decays much slower than the Poisson distribution, thus there are some vertices with very high degree. These vertices are the hubs and have role to keep the integrity and robustness of the network.
\end{enumerate}

Of the three reasons above, the existence of hubs is considered to be the most important and surprising finding because: 
\begin{enumerate}
\item Prior to the works of \cite{25}, it was very natural to think that in general real datasets have Poisson distribution family (including binomial, normal, and Gaussian distribution).
\item The existence of very large hubs implies that virtually there is no limit for vertices to create and receive new edges (this is the reason for scale-free term picked by Barab\'asi and Albert).
\item There should be some fundamental principles that govern the evolution of the scale-free networks. The principles are described as \emph{growth} and \emph{preferential attachment} (\citealp{25}), where the probability of receiving a new edge is proportional to the number of edges a vertex already has.
\end{enumerate}

In this paper, we study the preferential attachment mechanism in trading networks. By using the supply and demand principle, we show that the preferential attachment in trading networks is opposite to the corresponding mechanism in WWW network. Because the preferential attachment is the principle behind the formulation of link structure ranking algorithms like PageRank and HITS (see section \ref{sec3} for details), we will use the differences to define a ranking algorithm for trading networks. The proposed algorithm will be HITS-based because there are two type of transactions to be captured, sellings and buyings. In network term, sellings are equivalent to creating new outlinks and buyings are equivalent to receiving new inlinks (see section \ref{sec4} for details about resource flows). So out of these algorithms, only HITS which produces two type of scores, authority scores that correspond with inlinks (buyings) and hub scores that correspond with outlinks (sellings), can be extended to trading networks. And by using the same framework when deriving the proposed algorithm for trading networks, we present a new approach to accelerate HITS computation. The preliminary results can be found in (\citealp{54,52})

\section{Preferential attachment} \label{sec2}
\noindent The preferential attachment is a concept introduced by Yule in 1925 (\citealp{29}) and then is used to describe a class of mechanisms in which the probability of receiving a quantity is proportional to the number of that quantity the object already has. It has appeared in several fields under different names; in information science it is known as the \emph{cumulative advantage} (\citealp{30}), in sociology as the \emph{Matthew effect} (\citealp{31}), and in economics as the \emph{Gibrat principle} (\citealp{32}). The preferential attachment is long known to be the principle behind the power-law distribution exhibited by some real datasets, for example the distributions of wealth accumulation (\citealp{33}), the distribution of the number of species per genus (\citealp{29}), the distributions of word frequencies used in books and documents (\citealp{34,35}), and the number of collaborators in scientific collaboration networks (\citealp{26,9,14,15}) among others.

However, the presence of the preferential attachment in the network evolution doesn't always produce the power-law degree distribution. If there are some constraints in generating new edges, usually the degree distribution will not be following the power-law because there are not many vertices with very high number of degree. But usually it will not be following the Poisson distribution either. Instead it will have non power-law but still right-skewed degree distribution. For example, power grid and air traffic have exponential distributions, friendship networks have Gaussian distributions, and movie actors network has an exponentially truncated power-law distribution (\citealp{36}). 

\cite{14} provide a robust test to detect the preferential attachment in the network evolution by observing the change of degree $\Delta k$ as a function of $k$ ($\Delta k\propto k^{v}$) for every vertex over some time intervals (thus requiring dynamic data which is not always available). If the preferential attachment exists, $v$ will be bigger than $0$. For perfectly scale-free network, $v$ will be equal to $1$. And if the mechanism doesn't exist, $v$ will be equal to $0$ (note that actually \cite{14} use integral of the probability of receiving new edges, $\kappa(k)=\int_1^k\Pi(k')\mathrm{d}k'$ where $\Pi(k')\propto k^v$ to define the test). Different from usual simple test by only plotting $p_k$, this method can distinguish networks with preferential attachment from merely random graphs with power-law degree distributions (\citealp{37}). However, on many occasions it is usually sufficient to utilize $p_k$ distribution, and actually this is the most common approach used by the researchers to detect the presence of preferential attachment in the real networks (\citealp{4,6,7,8,9,13,24,25,26,27,28,30,31,34}).

\section{PageRank and HITS} \label{sec3}
\noindent Around the same time with the finding of the preferential attachment in the network evolution, two groups of researchers started to realize the role of link structure of WWW network in determining the values of the pages. Links in WWW network are the hyperlinks created by the site owners to point to other relevant pages, favorite pages, popular pages, or pages that contain useful information (these were especially true in the beginning of WWW era where most hyperlinks were created by human and link spammers were rare). So, the hyperlinks reflect the opinion about the values of the pages; the more valuable the pages, the more inlinks they have. Thus, the hyperlink structure can be utilized to distinguish important pages from less important ones. 

The first link structure ranking algorithm was proposed by Brin and Page (\citealp{39,38}), known as PageRank, a popularity measure based on hypothesis of a random surfer that is infinitely following the hyperlink structure of WWW network. In the long run, the proportion of time a random surfer spends on a page depends on the number of inlinks the page has and on the number of inlinks other pages that point to it have. This is intuitive because the number of inlinks of a page reflects its reachability from other pages. And because the proportion of time it spends on a page reflects the value of the page, the PageRank score of a page is proportional to the number of inlinks the page has and to the number of inlinks other pages that point to it have. On the other hand, because the hyperlinks are the opinions or the recommendations created by the site owners to other pages, the values of the recommendations should be dropped if there are too many of them on a page. Thus, the PageRank score of a page is inversely proportional to the number of outlinks other pages that point to it have. Mathematically, PageRank is defined with the following equation:
\begin{align}
pr_i=\sum_{j\in \mathcal{B}_i}\frac{pr_j}{\mathrm{outdeg}_j}
\label{eq1}
\end{align}
where $pr_i$ denotes PageRank score of page $i$, $\mathrm{outdeg}_i$ denotes outdegree of $i$, and $\mathcal{B}_i$ denotes set of pages that point to $i$.

The above equation is a circular statement: the score of a page depends on the scores of other pages that point to it, and in turn the scores of those pages depend on the scores of other pages that point to them. To solve it, usually iterative procedure is employed with each page is given an initial value (usually set to $1/N$, where $N$ is the number of pages).
\begin{align}
pr_i^{(k+1)}=\sum_{j\in \mathcal{B}_i}\frac{pr_j^{(k)}}{\mathrm{outdeg}_j},\quad k = 1,\ldots, K
\label{eq2}
\end{align}
where $K$ denotes the final iteration where the predefined criterion is satisfied.

To get a more compact form, Equation \eqref{eq2} is rewritten in the matrix form.
\begin{align}
\mathbf{pr}^{(k+1)T} = \mathbf{pr}^{(k)T}\mathbf{Do}^{-1}\mathbf{L}
\label{eq3}
\end{align}
where $\mathbf{Do}=\mathrm{diag}(\mathrm{outdeg}_1,\ldots,\mathrm{outdeg}_N)$, $\mathbf{pr}^{(k)T}$ denotes the $1\times N$ PageRank vector at iteration $k$, and $\mathbf{L}$ denotes the adjacency matrix induced from WWW network where $[\mathbf{L}]_{ij} = 1$ if there is a hyperlink from $i$ to $j$, and $0$ otherwise.

Equation \eqref{eq3} is the problem of finding the dominant eigenvector of $(\mathbf{Do}^{-1}\mathbf{L})^T$ by using the power method (\citealp{59}). By Markov chains theory, Equation \eqref{eq3} converges to a unique positive PageRank vector $\mathbf{pr}$ if\mbox{}f $\mathbf{Do}^{-1}\mathbf{L}$ is \emph{stochastic}, \emph{irreducible}, and \emph{aperiodic} (\citealp{40}). 

A matrix is stochastic if\mbox{}f there is no zero row and all the rows are normalized. So, the first adjustment is to modify $\mathbf{Do}^{-1}\mathbf{L}$ into a stochastic matrix. Let $\mathbf{d}$ be $N\times 1$ dangling vector where its $n^{th}$ ($n=1,2,\ldots N$) entry is $1$ if $n$ is a dangling page and $0$ otherwise, and $\mathbf{e}^T$ be all-one $1\times N$ vector. The stochastic version of $\mathbf{Do}^{-1}\mathbf{L}$ is $\mathbf{S} = \mathbf{Do}^{-1}\mathbf{L} + (1/N)\mathbf{de}^T$. 

A matrix is irreducible if\mbox{}f its directed graph is strongly connected; for every pair of vertices, there is at least one path connecting them. And a matrix is aperiodic if\mbox{}f there is only one principal eigenvalue on the spectral circle. The irreducibility and aperiodicity properties can be enforced by replacing all zero entries of $\mathbf{S}$ with small positive numbers. Thus, the stochastic, irreducible, and aperiodic version of $\mathbf{Do}^{-1}\mathbf{L}$ is $\mathbf{P} = \alpha\,\mathbf{S} + (1/N)(1-\alpha)\mathbf{ee}^T$, where $0<\alpha<1$ denotes a scalar that controls proportion of time the random surfer follows the hyperlinks as opposed to teleporting (usually set to $0.85$). And Equation \eqref{eq3} can be rewritten as:
\begin{align}
\mathbf{pr}^{(k+1)T} &= \mathbf{pr}^{(k)T}\mathbf{P} \nonumber \\
                    &= \alpha\mathbf{pr}^{(k)T}\mathbf{Do}^{-1}\mathbf{L} + (1/N)\Big(\alpha\mathbf{pr}^{(k)T}\mathbf{d} + 1-\alpha\Big)\mathbf{e}^T \label{eq4}
\end{align}

The second ranking algorithm, HITS (\emph{Hypertext Induced Topic Search}) was introduced by Kleinberg (\citealp{27}). Different from PageRank, HITS produces two metrics associated with every page, authority and hub. Authority scores determine pages' popularity and hub scores are used to find portal pages, pages that link to popular (thus useful) pages. 

HITS is defined with the following statement: \emph{authority score of a page is the sum of hub scores of others that point to it and hub score of a page is the sum of authority scores of others that are pointed to by it} (\citealp{27}). Like PageRank, this is also a circular statement, the authority scores depend on the hub scores and vice versa. To solve it, the following equation is used.
\begin{align}
a^{(k+1)}_{i} = \sum_{j\in \mathcal{B}_i} h^{(k)}_{j},\enspace\text{and} \enspace h^{(k+1)}_{i} = \sum_{j\in \mathcal{F}_i} a^{(k+1)}_{j}
\label{eq5}
\end{align}
where $a_{i}^{(k)}$ and $h_{i}^{(k)}$ denote the authority and hub score of page $i$ at iteration $k$, $\mathcal{B}_i$ denotes the set of pages that point to $i$, and $\mathcal{F}_i$ denotes the set of pages that are pointed to by $i$. In the matrix from, HITS formulation can be rewritten as:
\begin{align}
\mathbf{a}^{(k+1)T} = \mathbf{h}^{(k)T}\mathbf{L},
\enspace \text{and} \enspace
\mathbf{h}^{(k+1)T} = \mathbf{a}^{(k+1)T}\mathbf{L}^T
\label{eq6}
\end{align}
where $\mathbf{a}^{T}$ denotes 1$\times N$ authority vector and $\mathbf{h}^{T}$ denotes 1$\times N$ hub vector.

In HITS both authority matrix, $\mathbf{L}^{T}\mathbf{L}$ ($\mathbf{a}^{(k+1)T}=\mathbf{a}^{(k)T}\mathbf{L}^{T}\mathbf{L}$) and hub matrix, $\mathbf{LL}^{T}$ ($\mathbf{h}^{(k+1)T}=\mathbf{h}^{(k)T}\mathbf{L}\mathbf{L}^{T}$) are nonnegative. Thus by Perron theorem for nonnegative matrices, $\mathbf{a}^T$ and $\mathbf{h}^T$ exist but there is no guarantee of the uniqueness. To ensure the uniqueness, the authority and hub matrices must be modified into positive matrices (\citealp{55,40}). Let $\mathbf{\hat{A}}$ and $\mathbf{\hat{H}}$ be the positive version of the authority matrix and hub matrix respectively. We can define them as $\mathbf{\hat{A}} = \zeta\,\mathbf{L}^{T}\mathbf{L}+(1/N)(1-\zeta)\,\mathbf{e}\,\mathbf{e}^T$, and $\mathbf{\hat{H}} = \zeta\,\mathbf{L}\,\mathbf{L}^{T}+(1/N)(1-\zeta)\,\mathbf{e}\,\mathbf{e}^T$, where $0<\zeta<1$ denotes a constant that should be set near to $1$ to preserve the hyperlink structure information. Thus, unique and positive authority and hub vectors can be calculated by using $\mathbf{a}^{(k+1)T}=\mathbf{a}^{(k)T}\mathbf{\hat{A}}$ and $\mathbf{h}^{(k+1)T}=\mathbf{h}^{(k)T}\mathbf{\hat{H}}$.

\section{Link structure ranking algorithm for trading networks} \label{sec4}
\begin{figure}[t]
 \begin{center}
  \includegraphics[width=0.7\textwidth]{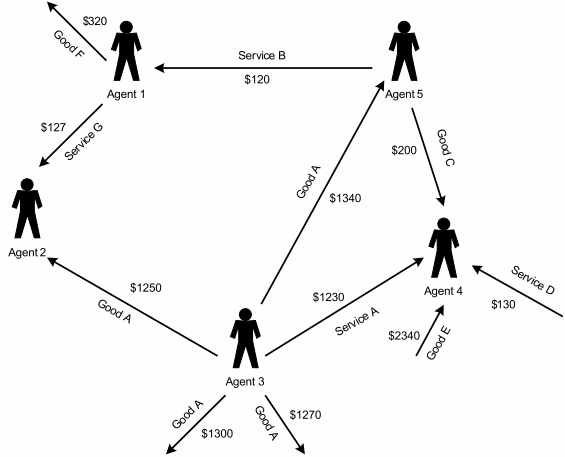}
  \caption{Labeled-link network model of the trading activities.}
  \label{fig2}
 \end{center}
\end{figure}

\noindent The trading activities are the exchanges of different goods and/or services (we will refer goods/services as resources for the rest of the paper) involving at least two agents. These activities can be modeled with labeled-link network where the vertices are the agents and the directed edges are the flows of the resources. Figure \ref{fig2} shows the network model of the trading activities. Note that actually the transactions are mutual; there are two opposite flows for each transaction, the flow of the resource and the flow of the payment. However because the price is a better unit of account in the market and generally is used to measure the quantity of the resources, each transaction can be described by only one directed edge, the flow of the resource weighted with the price. 

There are some differences between trading network and WWW network that are worth to be noted. \emph{First}, in trading network every vertex has at least one type of resource and a new edge is created when two vertices exchange their resources. Consequently, the amount of resources limits the number and the weight of edges a vertex can have. On the other hand, in WWW network the creation of edges is simply the creation of new hyperlinks on the web pages, so there is no resource needs to be allocated. \emph{Second}, different from trading network, the creation of edges in WWW network is not a mutual process; if page $A$ has a hyperlink to page $B$, it doesn't necessary that $B$ also has a hyperlink to $A$. \emph{Third}, every edge in trading network is labeled with resource description and is weighted with the price. On the other hand, the edges in WWW network are usually unlabeled and weighted with either $1$ or $0$\protect\footnote{There are some works that are devoted to the analysis of WWW network labels (the hypertexts). For example: \cite{42}, \cite{43}, and \cite{44}. But because the hyperlinks are the recommendations, they are alike, and in some cases can be ignored safely, including the calculations of PageRank and HITS. Conversely, in trading network the labels are the inherent information of the transactions that cannot be ignored at any cost}. And \emph{fourth}, while the purpose of edges creation in trading network is to maximize the transaction benefits, in WWW network is to get hyperlinks from popular pages.

The last difference is directly related to the preferential attachment mechanisms. Before we define the preferential attachment in trading network, we will enlist some assumptions that have to be taken in order to simplify the complex interactions among agents.
\begin{enumerate}
\item All transactions are carried out under \emph{ceteris paribus} condition. So the prices depend only on demands and supplies of the corresponding resources, not other substitute or complementary resources. 
\item The perfect market condition is met and the prices have already reached the equilibrium states.
\item The amount of resources owned by an agent is reflected in its buying and selling volumes of the corresponding resources.
\end{enumerate}

The first assumption allows us to form and analyze one network for each resource independently. The second assumption guarantees that resources availability is the main motivation in choosing business partners, not the price differences. And the third assumption allows us to estimate the resources availability for future transactions by using current and past buying and selling volumes of the corresponding resources, which is reflected in the weights of the inlinks and outlinks. 

Note that both first and second assumptions are very common in the trading network analysis and the economics in general. So, we will only discuss the reasons behind the last assumption. The last assumption is the heart of the proposed algorithm formulation because it allows us to (1) model the trading activities completely with the labeled-link network which is a standard model in graph theory, (2) relate the amount of resources owned by an agent to the weights of the corresponding inlinks and outlinks, and (3) define the preferential attachment in trading network by using the number of inlinks and outlinks (more specifically, total weights of those links) so that it can be compared to the preferential attachment in WWW network induced from the HITS formulation (see Figure \ref{fig3}), and in turn allowing us to formulate a ranking algorithm for trading network.

\subsection{Proposed algorithm formulation} \label{propalg}
In trading activities, there are costs associated with every transaction. Thus, every agent must implement an optimal preferential attachment strategy to maximize the benefits. In the real situation, every transaction conducted by an agent influences its financial states, including transactions from different resources. However, by assumption 1 we can isolate the influences and form one labeled-link network for each different resource. So, if there are $x$ type of resources traded among the agents, there will be $x$ labeled-link networks that can be analyzed separately. Then by assumption 2, each agent should buy (receive inlinks) from others with abundant resources, and should sell (create outlinks) to others that are lack of the resources. And by assumption 3, agents with abundant resources are the agents with many inlinks and agents that lack of the resources are the agents with many outlinks.

\begin{figure}[t]
 \begin{center}
 
  \subfigure[Trading network]{
   \includegraphics[width=0.35\textwidth]{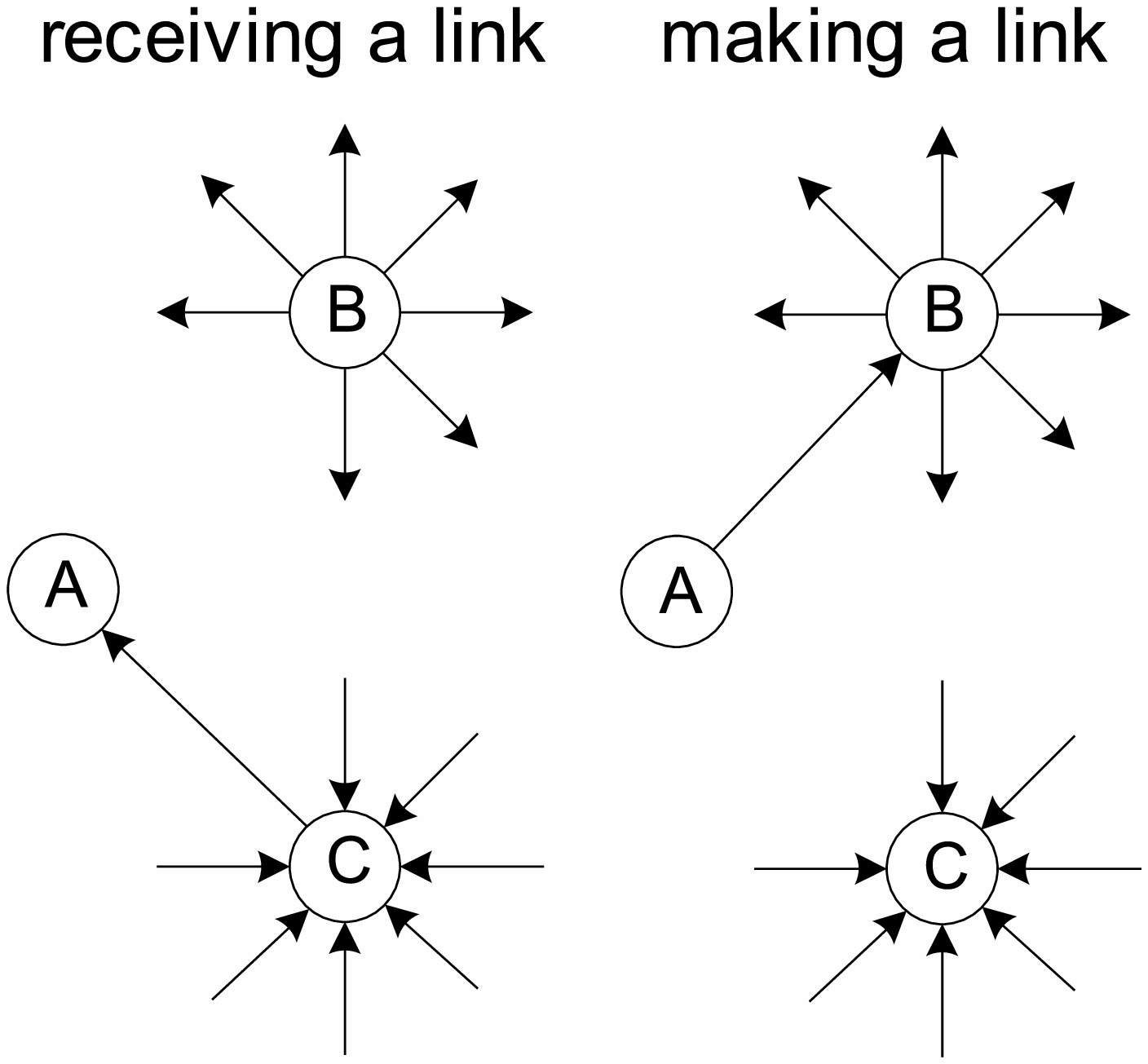}
   \label{tradingNetwork}
  }
  \subfigure[WWW network]{
   \includegraphics[width=0.35\textwidth]{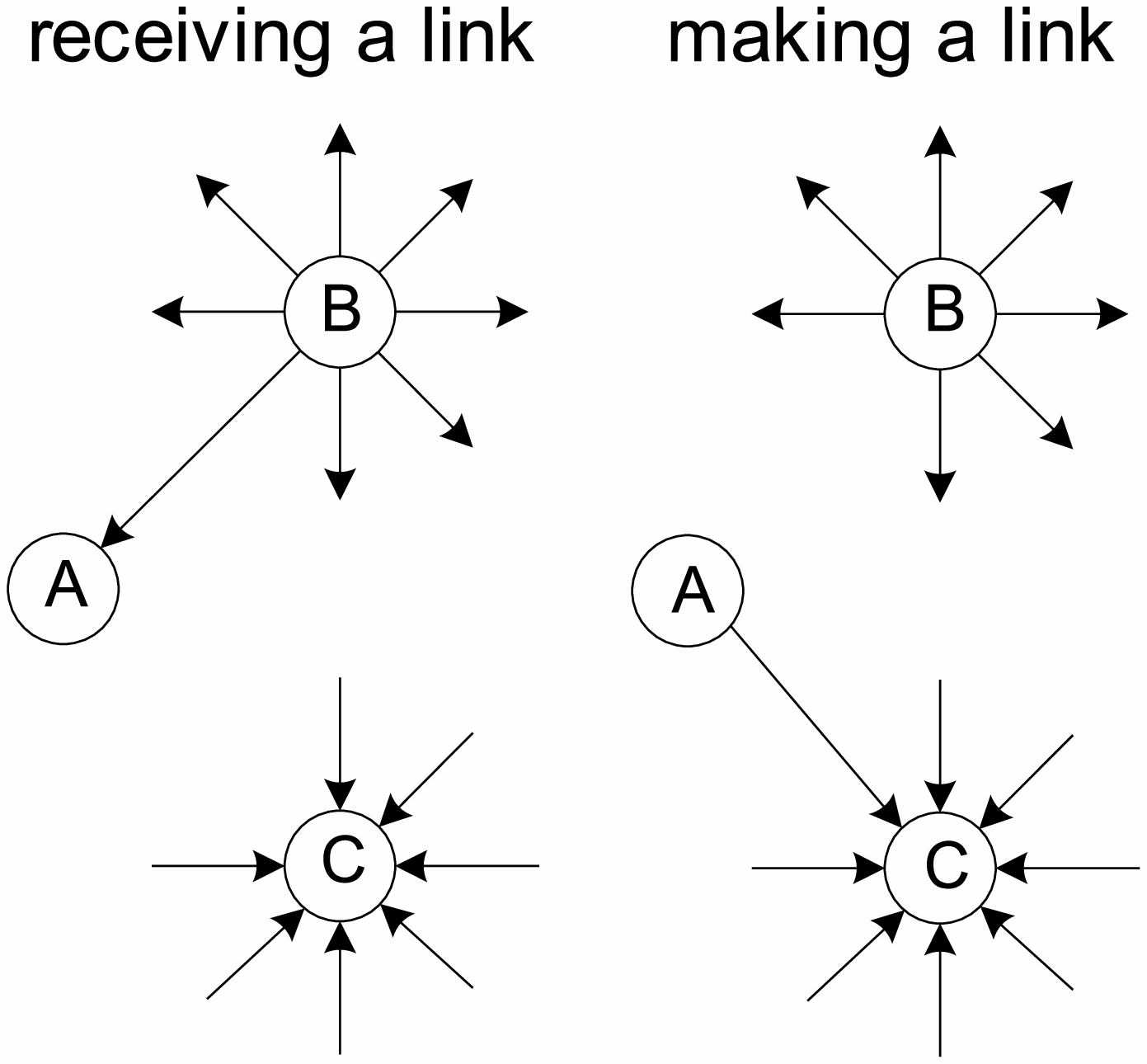}
   \label{WWWNetwork}
  }
  \caption{The preferential attachment mechanisms in trading network and WWW network (HITS's version).}
  \label{fig3}
 \end{center}
\end{figure}

Thus, we can define the preferential attachment in trading network with the following statement: \emph{an agent should receive new inlinks from others with many inlinks and should create new outlinks to others with many outlinks}. This statement is interesting because it resembles HITS's version of preferential attachment in WWW network. As discussed in section \ref{sec3}, in HITS good authorities (pages with many inlinks) are pointed to by good hubs (pages with many outlinks) and good hubs point to good authorities. Thus, HITS's version of preferential attachment is: \emph{a page should receive new inlinks from others with many outlinks and should create new outlinks to others with many inlinks}. Figure \ref{tradingNetwork} and \ref{WWWNetwork} show the preferential attachment in trading network and WWW network respectively. As we can see, the preferential attachment in trading network is opposite to the HITS's version of preferential attachment in WWW network, so it can be utilized to formulate the proposed algorithm.

The proposed algorithm is defined with the following statement: \emph{a vertex becomes more important if being pointed to by others with many inlinks and points to others with many outlinks}. This statement is derived directly from the preferential attachment in trading network defined above. And by comparing the preferential attachments in both networks (see Figure \ref{fig3}) and the HITS formulation (see Equation \ref{eq5}), the proposed algorithm can be written as:
\begin{align}
r^{(k+1)}_{i} &= \beta\sum_{j\in \mathcal{B}_i}r^{(k)}_{j}ca_j + (1-\beta)\sum_{j\in \mathcal{F}_i}r^{(k)}_{j}ch_j, \enspace \text{where} \label{eq7} 
\\
ca_{i} &= \frac{\text{indeg}_{i}}{\text{deg}_{i}}|\text{indeg}_{i} - \text{outdeg}_{i}|^{p_{i}}, \label{eq8} 
\\ 
ch_{i} &= \frac{\text{outdeg}_{i}}{\text{deg}_{i}}|\text{indeg}_{i} - \text{outdeg}_{i}|^{-p_{i}}, \enspace \text{and} \label{eq9}
\\
p_{i} &= \left\{
 \begin{array}{rl}
  1 \enspace & \text{if } \text{indeg}_{i} > \text{outdeg}_{i} \\
  -1 \enspace & \text{if } \text{indeg}_{i} < \text{outdeg}_{i} \\
  0 \enspace & \text{otherwise}
 \end{array} \right. \label{eq10}
\end{align}
where $r^{(k)}_{i}$ denotes ranking score of vertex $i$ at iteration $k$; indeg$_i$, outdeg$_i$, and deg$_i$ denote indegree, outdegree, and degree of $i$; and $0<\beta<1$ is a scalar used to determine which link is more important. If outlink (selling) is more important than inlink (buying), $\beta<0.5$; if inlink is more important than outlink, $\beta>0.5$; and $\beta=0.5$ otherwise.

The constants $ca$ and $ch$ are introduced to favour the preferential attachment. As shown in Equation \ref{eq8} and \ref{eq9}, $ca$ will be bigger for vertices with many inlinks, and $ch$ will be bigger for vertices with many outlinks. Thus, by Equation \ref{eq7}, vertices that are pointed to by others with many inlinks and point to others with many outlinks (following the preferential attachment) will have bigger scores than vertices that do the opposite (not following the preferential attachment). 

Note that the first term of the right hand part of Equation \ref{eq7}, $\sum_{j\in \mathcal{B}_i}r^{(k)}_{j}ca_j$, describes the fraction of scores a vertex receives from its inlinks, and the second term of the right hand part, $\sum_{j\in \mathcal{F}_i}r^{(k)}_{j}ch_j$, describes the fraction of scores a vertex receives from its outlinks. So, the first term can be defined as the authority part and the second term as the hub part.

The proposed algorithm will be represented in matrix to allow necessary adjustments be applied in order to ensure the convergence. Let $\mathbf{M} = \beta\mathbf{F}+(1-\beta)\mathbf{G}$, where $\mathbf{F}=\mathbf{K}\mathbf{D}^{-1}\mathbf{Di}\,\mathbf{L}$ be the authority part, and $\mathbf{G}=\mathbf{K}^{-1}\mathbf{D}^{-1}\mathbf{Do}\,\mathbf{L}^{T}$ be the hub part. Then, Equation \ref{eq7} can be rewritten as:
\begin{align}
\mathbf{r}^{(k+1)T} &= \mathbf{r}^{(k)T}\mathbf{M} \nonumber \\
&= \mathbf{r}^{(k)T}\Big(\beta\mathbf{K}\mathbf{D}^{-1}\mathbf{Di}\,\mathbf{L} + (1-\beta)\mathbf{K}^{-1}\mathbf{D}^{-1}\mathbf{Do}\,\mathbf{L}^{T}\Big)
\label{eq11}
\end{align}
where $\mathbf{L}$ denotes the induced adjacency matrix, $\mathbf{r}^{(k)T}$ denotes $1\times N$ ranking vector at interation $k$, $\mathbf{Di}=\mathrm{diag}(\mathrm{indeg}_1,\ldots,\mathrm{indeg}_N)$, $\mathbf{Do}=\mathrm{diag}(\mathrm{outdeg}_1,\ldots,\mathrm{outdeg}_N)$, $\mathbf{D}=\mathbf{Di}\,+\, \mathbf{Do}$, and $\mathbf{K}$ is a diagonal matrix where $[\mathbf{K}]_{ii}=|(\mathbf{Di}-\mathbf{Do})_{ii}|^{p_i}$. Note that different from WWW network, in trading network entries of $\mathbf{L}$ are the weights of the corresponding links which are usually nonnegative real numbers. 

As shown in Equation \ref{eq11}, $\mathbf{M}$ has no zero row, but is not a stochastic matrix because the rows are not normalized. Therefore the stochasticity adjustment is required. Let $\mathbf{N}$ be a diagonal matrix where $[\mathbf{N}]_{ii}=\sum_{j\in \mathcal{V}}\mathbf{M}_{ij}$ ($\mathcal{V}$ denotes the set of all vertices in the network), the stochastic version of $\mathbf{M}$ can be written as $\mathbf{\overline{M}}=\mathbf{N}^{-1}\mathbf{M}$. And the irreducibility and aperiodicity adjustments can be done by replacing all zero entries of $\mathbf{\overline{M}}$ with small positive numbers: $\mathbf{R}=\zeta\,\mathbf{\overline{M}}+(1/N)(1-\zeta)\mathbf{ee}^T$, where $0<\zeta<1$ is equivalent to $\alpha$ in PageRank and should be set near to $1$. Thus, the proposed algorithm can be rewritten as:
\begin{align}
\mathbf{r}^{(k+1)T} = \mathbf{r}^{(k)T}\mathbf{R}
\label{eq12}
\end{align}
As we can see, $\mathbf{R}$ is identical to $\mathbf{P}$ in Equation \ref{eq4}, and by choosing a positive initial vector (for example $\mathbf{r}^{(k=1)T}=(1/N)\mathbf{e}^T$) the Equation \ref{eq12} is guaranteed to converge to a unique positive ranking vector $\mathbf{r}^{(K)T}$ (\citealp{55}).

The proposed algorithm only accommodates the flowing resources. If we have data about the amount of resources owned by the agents which is not from the transactions, for example natural resources like gas, oil, coal, gold, etc (we will refer these as reserved resources for the rest of the paper), this information can also be included in the final scores. Let $\mathbf{u}^T$ be $1\times N$ vector where $[\mathbf{u}]_i$ corresponds to the amount of the reserved resource of agent $i$. Then the final ranking vector can be written as: $\mathbf{\hat{r}}^T=c\,\mathbf{r}^T+(1-c)\mathbf{\overline{u}}^T$, where $\mathbf{\overline{u}}$ is the normalized version of $\mathbf{u}$, and $0<c<1$ is a control parameter that determines which vector is more important. 

We can also introduce a scaling constant similar to the work of \cite{16} associated with every agent to the final score to describe its competitiveness. These constants can be used not only to favour the competitive agents, but also to handle some issues related to the trading activities like reliability and trust issues.

Occasionally, agents' scores as the buyers and/or the sellers are more desirable than the overall scores. By inspecting Figure \ref{tradingNetwork} and Equation \eqref{eq7}, ranking vector as the buyers, $\mathbf{b}^T$, and as the sellers, $\mathbf{s}^T$, can be written as:
\begin{align}
\mathbf{b}^{(k+1)T} = \mathbf{b}^{(k)T}\mathbf{Ca}\,\mathbf{L},
\enspace \text{and} \enspace
\mathbf{s}^{(k+1)T} = \mathbf{s}^{(k)T}\mathbf{Ch}\,\mathbf{L}^T
\label{eq13}
\end{align}
where $\mathbf{Ca} = \text{diag}(ca_1, \ldots, ca_N)$, and $\mathbf{Ch} = \text{diag}(ch_1, \ldots, ch_N)$.

\subsection{Experimental results}
\noindent We will examine the proposed algorithm performance by using international trading datasets from the United Nations (\citealp{47,46}). There are several good reasons in choosing these datasets. \emph{First}, the size of the networks are small compared to other datasets like online auction networks, therefore the errors produced in each iteration can be minimized. \emph{Second}, the classification of products is clear, so the adjacency matrix for every product can be easily constructed. And \emph{third}, the prices of the products in the same category are almost the same, complying with the second assumption.

As stated earlier $\mathbf{R}$ is stochastic, irreducible, and aperiodic. Thus, the power method applied to Equation \eqref{eq12} is guaranteed to converge to a unique positive ranking vector $\mathbf{r}^{(K)T}$ for any positive starting vector. Therefore, the question left is ``will it converge to something that makes sense in the context of measuring the degree of importance of agents in trading network''. We will answer this question by calculating the similarity between vector of our proposed algorithm $\mathbf{r}$, and standard measure, vector of total export and import $\mathbf{t}$. This vector is chosen as the standard measure not only because it is the simplest and common way in measuring the degree of importance, but also because the most active agents are usually the most connected ones which are conventionally considered to be the most important vertices in the graph theory. And as the similarity measures, cosine criterion $\cos\theta$ and Spearman rank order correlation coefficient $\rho$ will be used.
\begin{align}
\cos\theta = \frac{\mathbf{r}\cdot \mathbf{t}}{\Vert \mathbf{r}\Vert_2 \Vert \mathbf{t}\Vert_2}, \enspace \mathrm{and} \enspace
\rho = 1-\frac{6\sum_{i=1}^{N}([o(\mathbf{r})]_i-[o(\mathbf{t})]_i)^2}{N(N^2-1)}
\label{eq14}
\end{align}
where $\Vert\ast \Vert_2$ denotes 2-norm of vector $\ast$, and $o(\ast)$ denotes the ordering induced from vector $\ast$. For example, if $\ast=[0.3397, 0.1819, 0.3328]$, then $o(\ast)=[1,3,2]$. Thus, while the cosine criterion measures the distance between two vectors, the Spearman correlation measures the similarity between orderings induced from the vectors.
\begin{table*}[b]
 \begin{center}
  \caption{The performances of the proposed algorithm.}
   \footnotesize{\begin{tabular}[t]{|l|r|r|r|r|r|c|c|}
    \hline
Data & \#Vert. & \#Edg. & \multicolumn{3}{c|}{\#Iterations} & \multicolumn{2}{c|}{Similarity} \\ \cline{4-8}
 & & & HITS & PR & Prop. Alg. & $\cos\theta$ & $\rho$ \\
\hline
Steel products     & 97 & 2627 & 26 & 54 &  42 & 0.862 & 0.874 \\
Ethylene           & 43 &  169 &  7 & 44 &  54 & 0.849 & 0.916 \\
Propylene          & 38 &  144 & 10 & 40 & 143 & 0.974 & 0.905 \\
Sodium             & 49 &  268 & 11 & 53 & 143 & 0.808 & 0.850 \\
Hydrogen peroxide  & 47 &  261 & 51 & 61 &  99 & 0.752 & 0.902 \\
Carbon             & 51 &  535 & 22 & 37 &  65 & 0.912 & 0.929 \\
Radio-active       & 53 &  717 & 25 & 23 &  26 & 0.884 & 0.927 \\
Plastics           & 53 & 1410 & 20 & 37 &  39 & 0.985 & 0.968 \\
Medicinal products & 53 & 1504 &  9 & 18 &  14 & 0.989 & 0.965 \\
\hline
Average            & 54 &  848 & 20 & 41 &  69 & 0.891 & 0.915 \\
\hline
   \end{tabular}}
   \label{table1}
 \end{center}
\end{table*}
\begin{table*}[t]
 \begin{center}
  \caption{Top ten countries in hydrogen peroxide trading.}
   \footnotesize{\begin{tabular}[t]{|l|c|l|c|}
    \hline
\multicolumn{2}{|c|}{Ordered by stand. meas.} & \multicolumn{2}{c|}{Ordered by prop. alg.} \\
\hline
Country & Score & Country & Score \\
\hline
Netherlands     & 0.132290 & Japan           & 0.172970 \\
Canada          & 0.095014 & Norway          & 0.123360 \\
United States   & 0.088694 & Netherlands     & 0.114200 \\
Moldova         & 0.065088 & Canada          & 0.082261 \\
Austria         & 0.059850 & Turkey          & 0.053170 \\
China           & 0.054194 & United States   & 0.047059 \\
Japan           & 0.048676 & Rep. Korea      & 0.043684 \\
Italy           & 0.045744 & Moldova         & 0.038344 \\
Colombia        & 0.037772 & China           & 0.036916 \\
Turkey          & 0.037353 & Thailand        & 0.034545 \\
\hline
   \end{tabular}}
   \label{table2}
 \end{center}
\end{table*}
\begin{table*}[t]
 \begin{center}
  \caption{Top ten countries in medicinal products trading.}
   \footnotesize{\begin{tabular}[t]{|l|c|l|c|}
    \hline
\multicolumn{2}{|c|}{Ordered by stand. meas.} & \multicolumn{2}{c|}{Ordered by prop. alg.} \\
\hline
Country & Score & Country & Score \\
\hline
Germany        & 0.133530 & Germany        & 0.139490 \\
United States  & 0.114520 & United Kingdom & 0.107270 \\
United Kingdom & 0.096001 & United States  & 0.098509 \\
France         & 0.092408 & Switzerland    & 0.095938 \\
Switzerland    & 0.083244 & France         & 0.085463 \\
Italy          & 0.067707 & Italy          & 0.064711 \\
Belg-Luxemb.   & 0.056696 & Belg-Luxemb.   & 0.051169 \\
Netherlands    & 0.051564 & Netherlands    & 0.047270 \\
Japan          & 0.049308 & Ireland        & 0.043663 \\
Sweden         & 0.033573 & Sweden         & 0.041134 \\
\hline
   \end{tabular}}
   \label{table3}
 \end{center}
\end{table*}

To get insight about the computational performance, the number of iterations required by the proposed algorithm to achieve the same residual level will be compared to the results of PageRank and HITS. In the experiments, the residual level is set to $10^{-8}$ and $\beta$ is set to $0.5$. The number of iterations is chosen instead of computational time because the sizes of trading networks are very small, so the power method produces negligible computational time. Table \ref{table1} gives summary of the results, and Table \ref{table2} and \ref{table3} show lists of top ten countries in hydrogen peroxide trading (the least similar to the standard measure in the cosine criterion) and medicinal products (the most similar to the standard measure in the cosine criterion).

As shown in Table \ref{table1}, the proposed algorithm takes more iteration steps to converge. But because trading networks are usually much smaller than WWW network, this is unlikely to become a problem (the computational times of these three algorithms are practically zero). And the similarity measures both in the cosine criterion and the Spearman correlation give promising results with average around 89\% and 91\% respectively. This high similarities are also confirmed by the top ten countries shown in Table \ref{table2} and \ref{table3}. Thus, it can be conferred that the proposed algorithm gives meaningful results in measuring the degree of importance of vertices in trading networks.

However, an important issue arises concerning the usefulness of the proposed algorithm. If the total volumes can describe the degree of importance, one can argue about the meaning of using the proposed algorithm which is clearly much more expensive to compute. Before answering this question, we should make clear that in general the problem of assigning the degree of importance to vertices in a graph doesn't have correct solution. Rather, the ``correct'' issue is how to find the useful solution. This issue has been extensively studied in WWW network where there are numerous methods which can roughly be classified into query-dependent scores and query-independent scores. For example content scores are query-dependent and PageRank is query-independent. And if the user satisfaction is considered to be the usefulness standard, PageRank seems to be more useful than HITS. 

Hence, the main purpose of the proposed algorithm is to present a new method to compute ranking scores in trading networks which will become crucial if the problem involving finding the most important and relevant users in a large trading network like online auction network (this is the recommendation problem which arises as one of the most important problem in the computer science researches (\citealp{57})). And because the proposed algorithm uses the network structure, an uncaptured information in the total volumes method, the amount of the resource is not only the factor, the link structure information is also important in determining the final scores. Thus in the proposed algorithm's viewpoint, a well connected vertex which can be considered an important vertex in the graph term are more favourable than a less connected vertex with the same amount of resource.

\section{Acceleration method for HITS} \label{sec5}
\noindent As shown in Figure \ref{fig3} the preferential attachment in WWW network induced from the HITS definition is opposite to the preferential attachment in trading network. Therefore, the same framework when deriving the ranking algorithm for trading network can be applied back to HITS. To derive the modified HITS formulation, we first discuss Equation \eqref{eq13} because it separates the ranking vector $\mathbf{r}$ into buying vector $\mathbf{b}$ and selling vector $\mathbf{s}$, so it is in the same shape with the HITS formulation in Equation \eqref{eq6}. By comparing the preferential attachment in trading network in Figure \ref{tradingNetwork} with Equation \eqref{eq13}, we can get insight about the relationship between the preferential attachment and the buying and selling vectors.

\subsection{Modified HITS formulation} \label{mhits}
\noindent As shown in the left hand side of Figure \ref{tradingNetwork}, an agent prefers other with many inlinks when receiving a new inlink. And in the first part of Equation \eqref{eq13}, ranking score of an agent as a buyer is the sum of ranking scores of others as buyers weighted with $ca$ of the corresponding agents from which it receives the resources. Because $ca$ is bigger if an agent has many inlinks than outlinks, the first part of Equation \eqref{eq13} says an agent should receive new inlinks from others with many inlinks, which is identical to the preferential attachment shown in the left hand side of Figure \ref{tradingNetwork}. 

This is also true for the selling part (right hand side of Figure \ref{tradingNetwork}); an agent prefers other with many outlinks when creating a new outlink. And in the second part of Equation \eqref{eq13}, ranking score of an agent as a seller is the sum of ranking scores of others as sellers weighted with $ch$ of the corresponding agents to which it delivers the resources. Because $ch$ is bigger if an agent has many outlinks, second part of Equation \eqref{eq13} says an agent should creates new outlinks to others with many outlinks, which is identical to the preferential attachment shown in the right hand side of Figure \ref{tradingNetwork}. Thus, it is clear that the preferential attachments in Figure \ref{fig3} can be utilized directly to the formulation of the ranking algorithms.

We will use this connection to define the modified HITS. As shown in the left hand side of Figure \ref{WWWNetwork}, a page prefers other with many outlinks when receving a new inlink. Because in WWW network inlink corresponds with authority concept and outlink corresponds with hub concept, this preferential attachment implies that authority score of a page is the sum of hub scores of others that point to it weighted with $ch$ of the corresponding pages. And in the right hand side of Figure \ref{WWWNetwork}, a page prefers other with many inlinks when creating a new outlink. Consequently, hub score of a page is the sum of authority scores of others that are pointed to by it weighted with $ca$ of the corresponding pages. Thus, the proposed algorithm can be written as:
\begin{align}
a^{(k+1)}_{i} = \sum_{j\in \mathcal{B}_i} h^{(k)}_{j}ch_{j},\enspace\text{and} \enspace h^{(k+1)}_{i} = \sum_{j\in \mathcal{F}_i} a^{(k+1)}_{j}ca_{j}
\label{eq15}
\end{align}
And in the matrix form:
\begin{align}
\mathbf{a}^{(k+1)T} = \mathbf{h}^{(k)T}\mathbf{Ch}\,\mathbf{L},
\enspace \text{and} \enspace
\mathbf{h}^{(k+1)T} = \mathbf{a}^{(k+1)T}\mathbf{Ca}\,\mathbf{L}^T
\label{eq16}
\end{align}

As shown in the Equation \eqref{eq15}, the proposed algorithm is HITS with the introduction of two constants to every page. Because $ca$ is bigger for an authoritative page and $ch$ is bigger for a hubby page, the pages that follow the preferential attachment will collect their scores faster as the iterations progress under the proposed algorithm than under HITS. Thus, it can be expected that the proposed algorithm will converge faster in the datasets that are following the preferential attachment. As shown in the previous works (\citealp{6,5,48}), WWW network does have power-law degree distributions for both inlinks and outlinks, so the preferential attachment exists.
\begin{figure}[t]
 \begin{center}
  \includegraphics[width=0.6\textwidth]{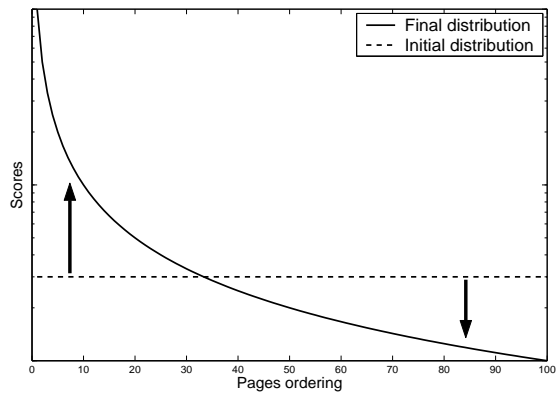}
  \caption{The distances between initial and final distributions.}
  \label{fig4}
 \end{center}
\end{figure}

Usually, a uniform distribution is used as the starting vector (\citealp{40}). Thus, the distances between initial and final scores are not uniform. For some very authoritative and hubby pages, it takes more iterations to reach the final scores. This is also true for pages that have very low final authority or hub scores. Figure \ref{fig4} describes such condition; the distances between initial and final scores of the pages that ordered in the top and bottom are greater than the pages in the middle positions. Because the authority and hub scores are proportional to $ca$ and $ch$\protect\footnote{As shown in (\citealp{50,49}), authority (hub) scores are proportional to the number of inlinks (outlinks), and by definition $ca$ ($ch$) values are proportional to the number of inlinks (outlinks). Thus the authority and hub scores are proportional to the $ca$ and $ch$ respectively.}, the distances between final and initial authority and hub scores are proportional to $ca$ and $ch$ respectively. Thus, the pages ordered in the top (bottom) will reach the stationary values faster under the proposed algorithm due to the bigger (smaller) $ca$ and $ch$. 

\subsection{Experimental results}
\noindent Due to the limited space, we only present the experimental results and analysis briefly. More detailed discussions can be found in (\citealp{56}).

There are six datasets used in the experiments that consist of around 10 thousands to 225 thousands pages with average degree from 4 to 47. Except wikipedia (\citealp{51}), all datasets were crawled by using our crawling system (\citealp{52}). All datasets, but britannica, have a typical WWW network average degree, around 4 to 15 (\citealp{40,53}). Table \ref{table4} summarizes the datasets where AD denotes the average degree.
\begin{table*}[t]
 \begin{center}
  \caption{Datasets summary.}
   \footnotesize{\begin{tabular}[t]{|l|l|r|r|r|}
    \hline
Data & Crawled & \#Pages & \#Links & AD\\
\hline
britannica   & 09/2008    & 21104   & 994554  & 47.1 \\
jobs         & 12/2008    & 16056   & 187957  & 11.7 \\
opera        & 12/2008    & 49749   & 437748  & 8.8  \\
scholarpedia & 06/2008    & 74243   & 1077781 & 14.5 \\
stanford     & 12/2008    & 225441  & 2196441 & 9.7  \\
wikipedia    & 09/2006    & 10431   & 46152   & 4.4  \\
\hline
\multicolumn{2}{|c|}{Average}  & 66170   & 46152   & 16   \\
\hline
   \end{tabular}}
   \label{table4}
 \end{center}
\end{table*}

\begin{figure}[t]
 \begin{center}
 
  \subfigure[http://www.britannica.com]{
   \includegraphics[width=0.4\textwidth]{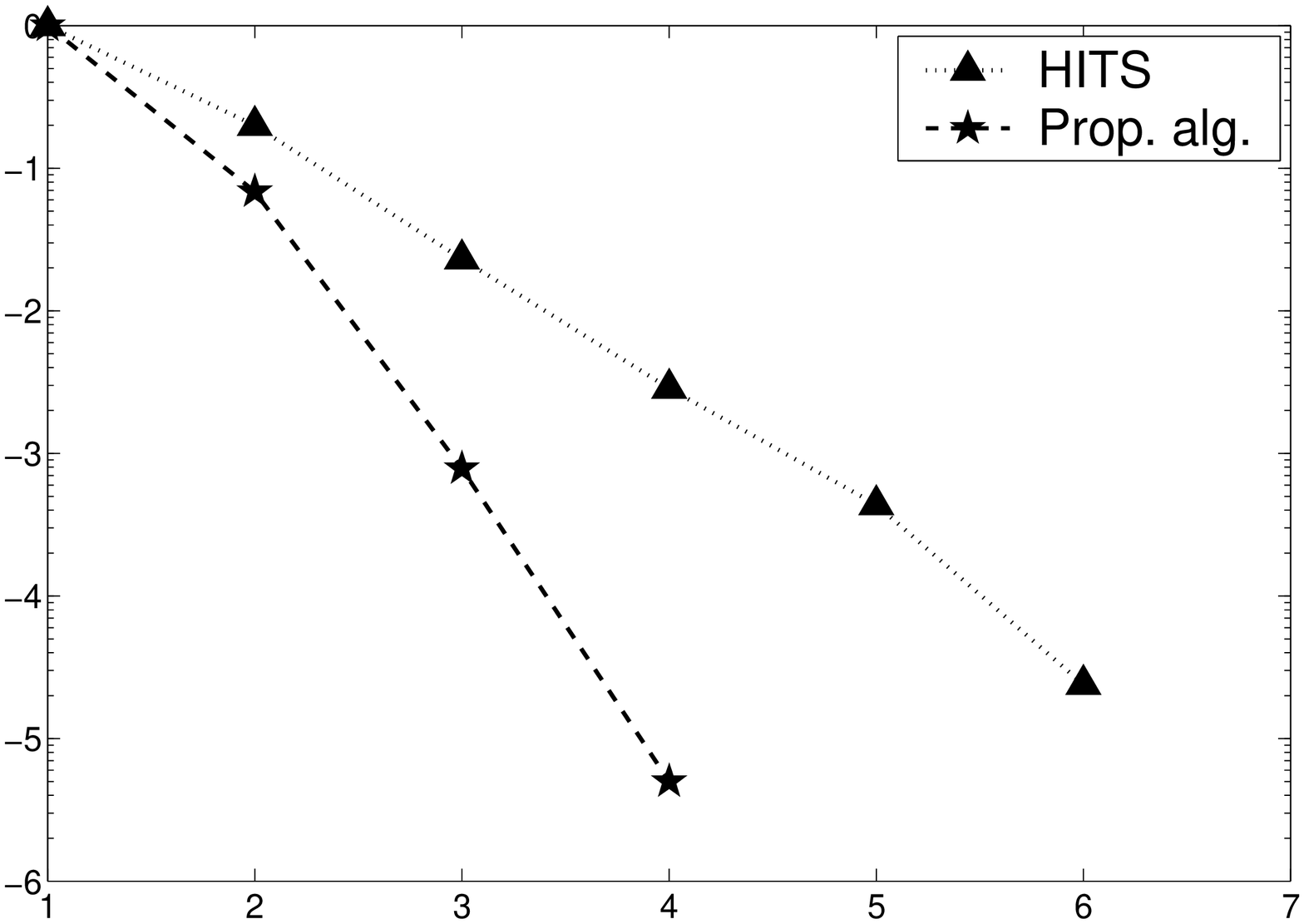}
   \label{britannica}
  }
  \subfigure[http://www.jobs.ac.uk]{
   \includegraphics[width=0.4\textwidth]{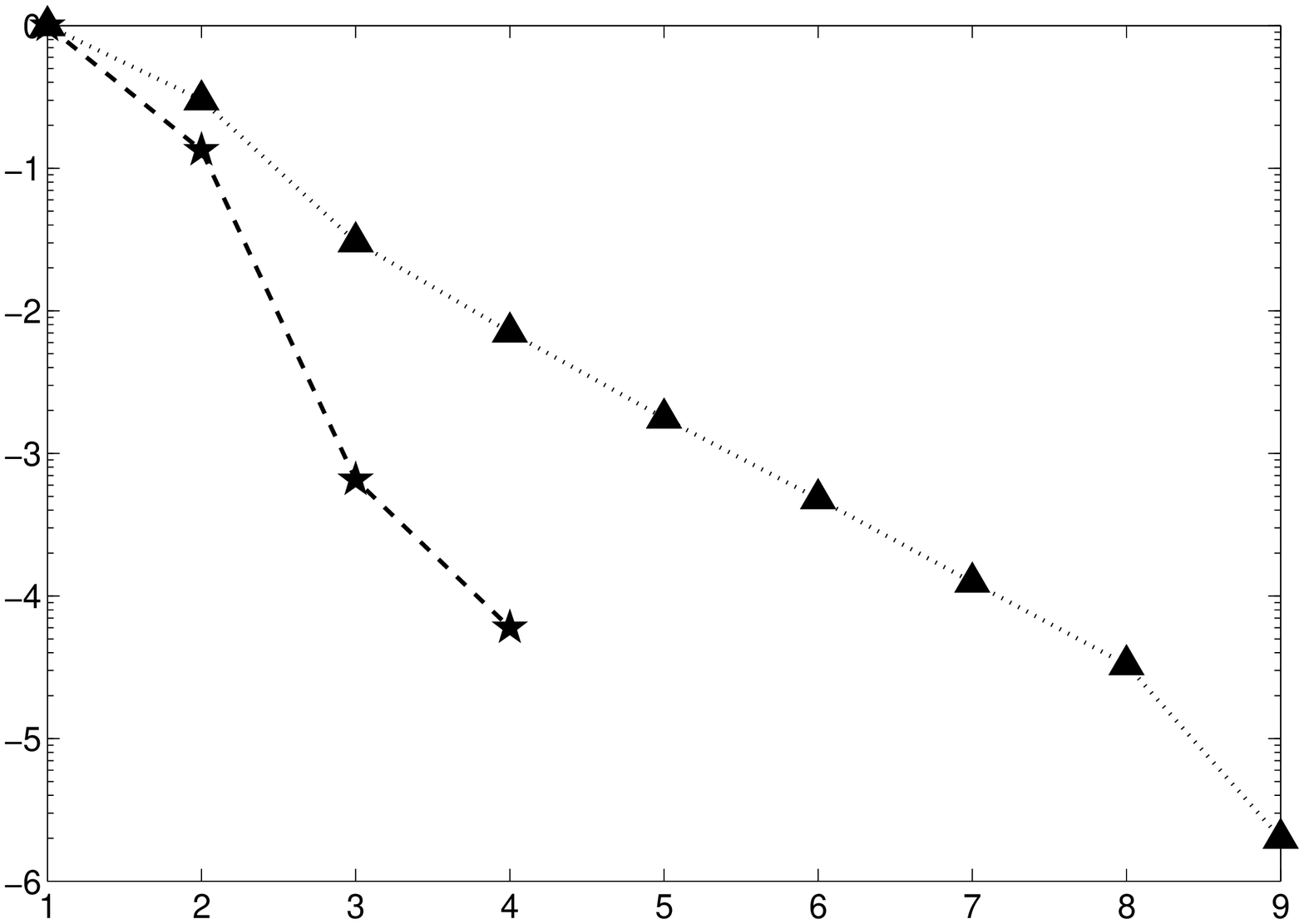}
   \label{jobs}
  }
  \\
  \subfigure[http://www.opera.com]{
   \includegraphics[width=0.4\textwidth]{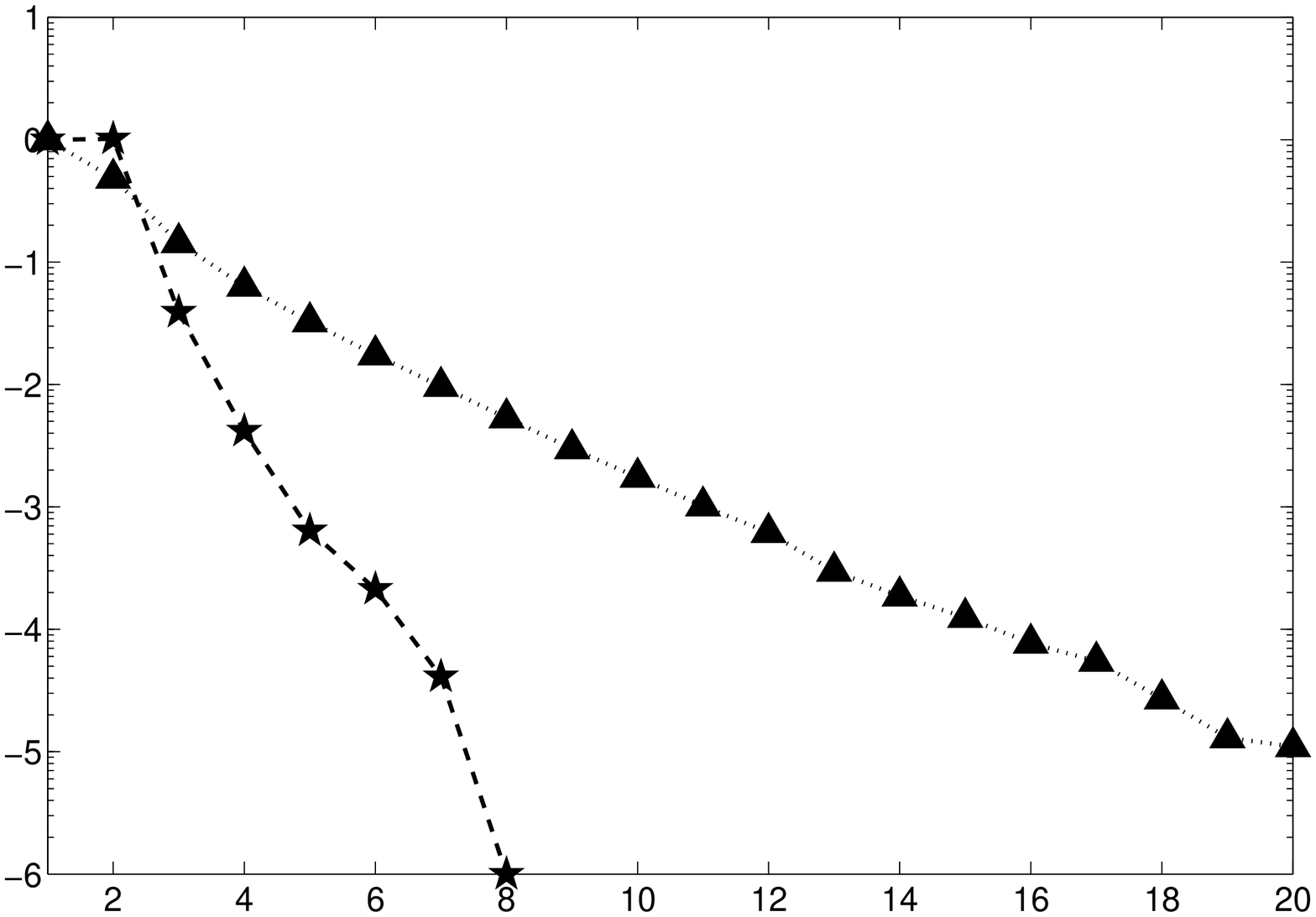}
   \label{opera}
  }
  \subfigure[http://www.scholarpedia.org]{
   \includegraphics[width=0.4\textwidth]{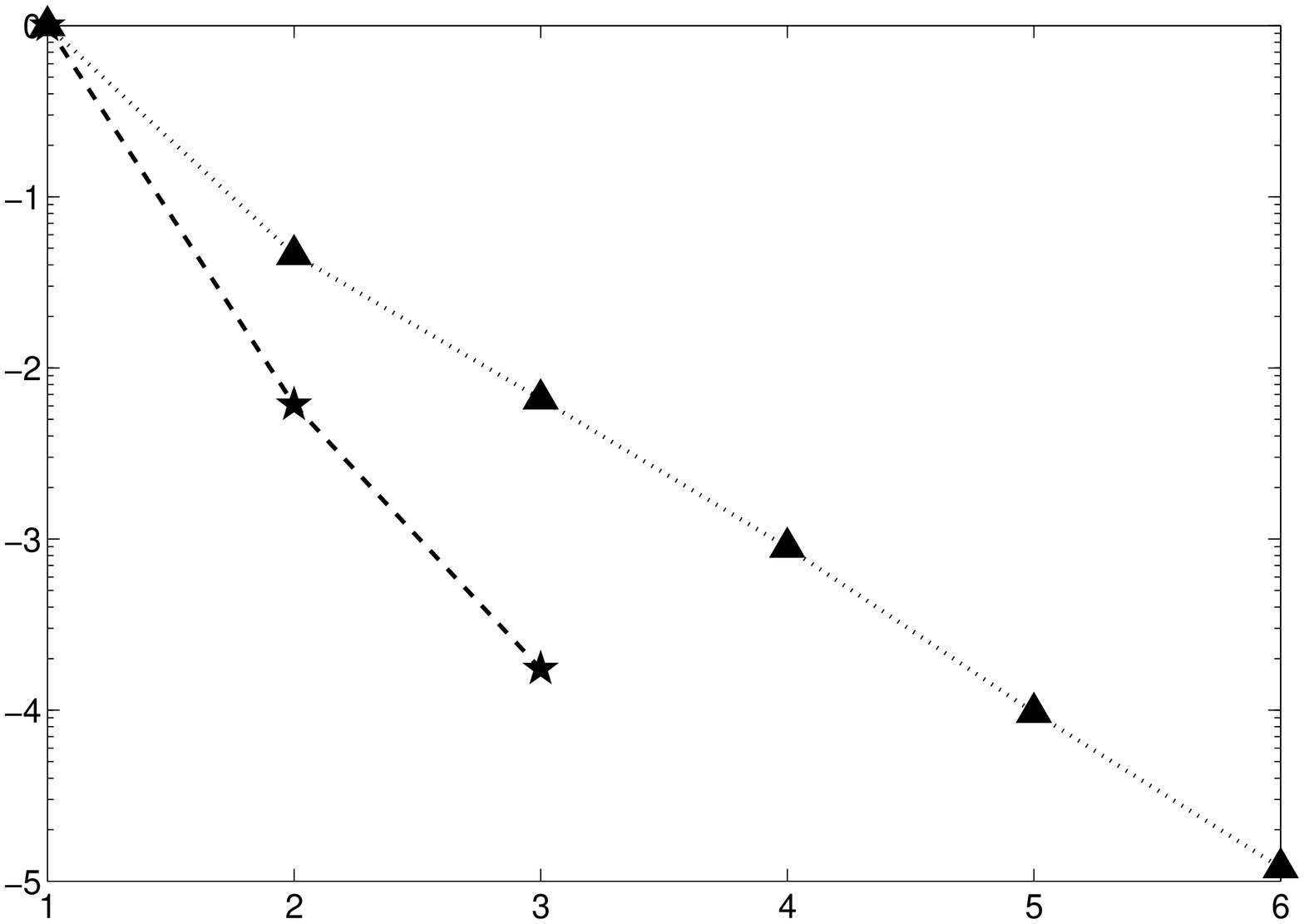}
   \label{scholarpedia}
  }
  \\
  \subfigure[http://www.stanford.edu]{
   \includegraphics[width=0.4\textwidth]{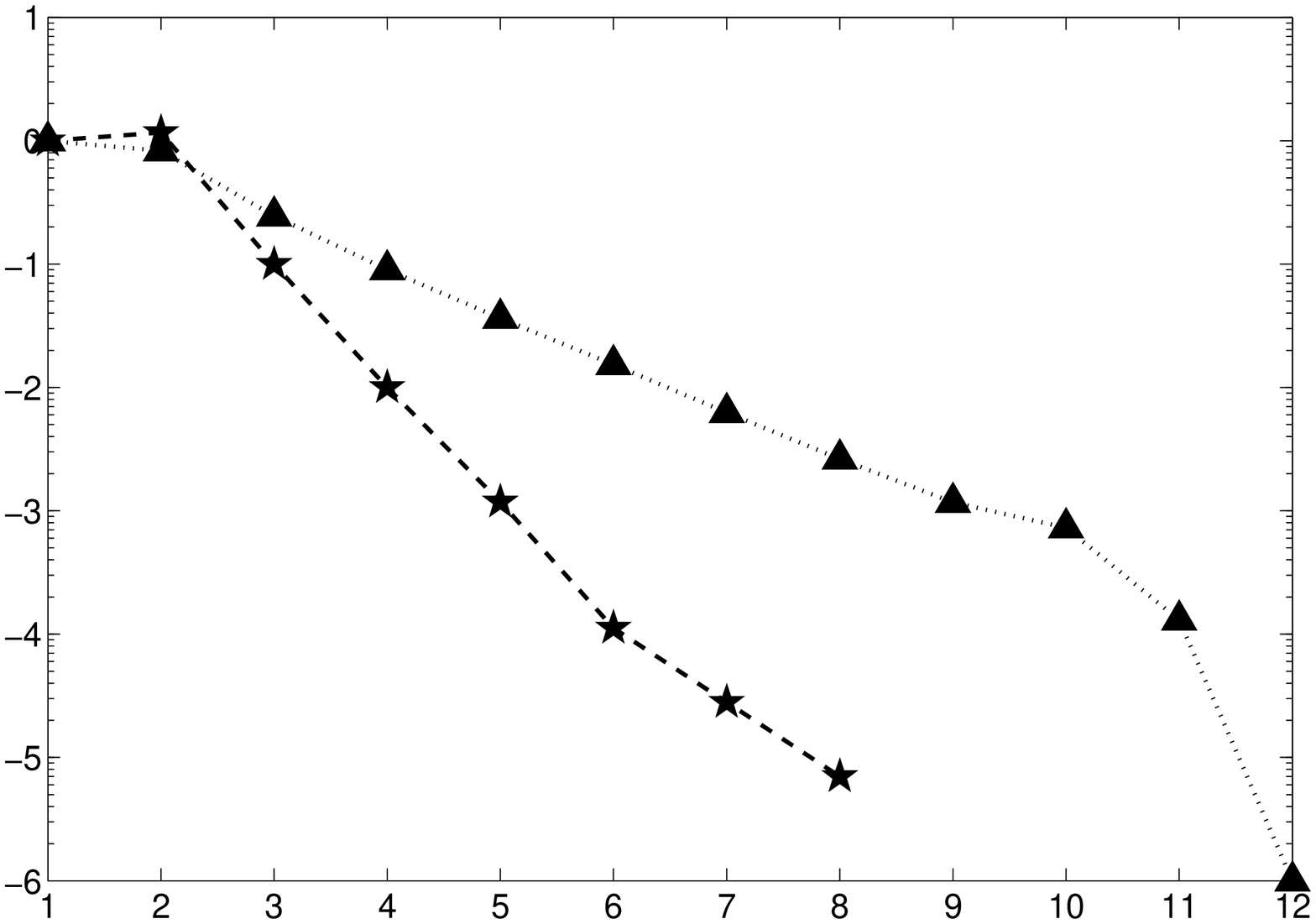}
   \label{stanford}
  }
  \subfigure[http://www.wikipedia.org]{
   \includegraphics[width=0.4\textwidth]{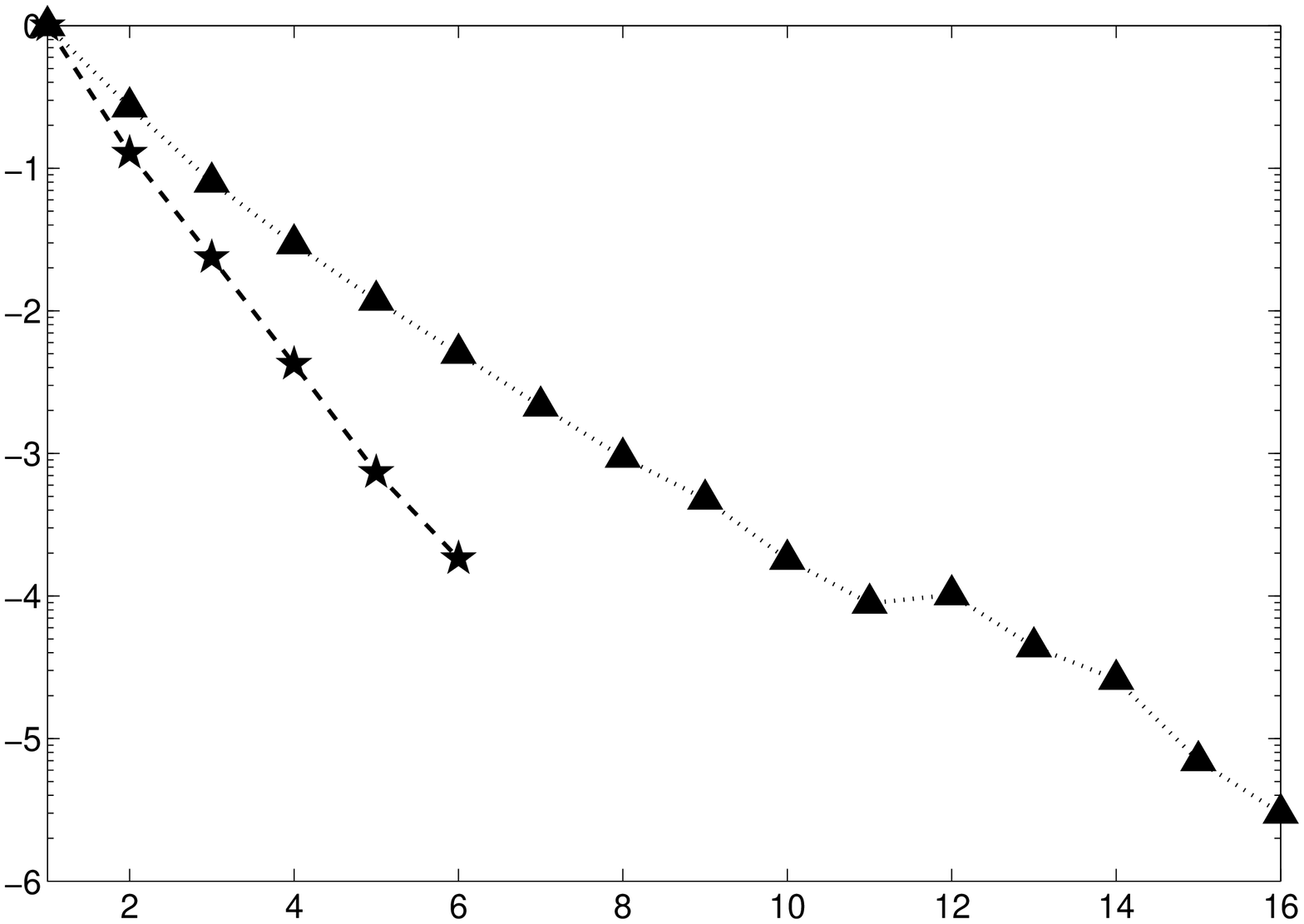}
   \label{wiki}
  }
  \caption{Convergence rates comparison. Note that x-axis is the number of iterations and y-axis is the residual in log scale.}
  \label{fig5}
 \end{center}
\end{figure}

\begin{figure}[tb]
 \begin{center}
  \includegraphics[width=0.5\textwidth]{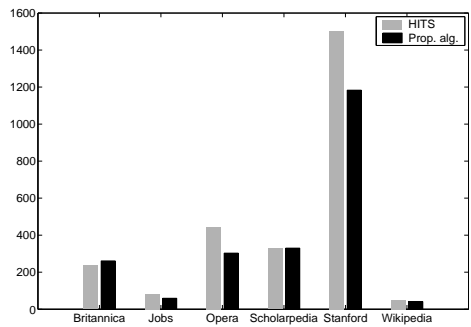}
  \caption{Processing times (in second) to reach the same corresponding residual levels.}
  \label{fig6}
 \end{center}
\end{figure}

The experiments are conducted by using a notebook with 1.86 GHz Intel Processor and 2 GB RAM. The codes are written in python by extensively using database to store lists of adjacency matrices, score vectors, and other related data. Figure \ref{fig5} shows the convergence rates and Figure \ref{fig6} shows processing times to achieve the same corresponding residual levels. Note that the uniform starting vectors are used for all datasets, and $ca$ and $ch$ computations have already been included in the processing times.

As shown in Figure \ref{fig5} and \ref{fig6}, the proposed algorithm in general can give improvements to both convergence rates and processing times. While in the processing times there are still some cases where the proposed algorithm cannot do better than HITS, in the convergence rates the proposed algorithm performs better than HITS in all cases. Table \ref{table5} gives examples of top ten pages returned by HITS and the proposed algorithm with query ``programming'' for wikipedia dataset. Note that for brevity only file names are displayed. To get full URLs, each name has to be prefixed with ``http://en.wikipedia.org/wiki/''.

\begin{table*}[tb]
 \begin{center}
   \caption{Top 10 results with query ``programming'' for wikipedia dataset.}
   \footnotesize{
    \begin{tabular}{|r|l|l|}
     \hline
     No. & HITS & Prop.~Alg. \\
     \hline
     1 & \verb|Programming_language| & \verb|Programming_language| \\
     2 & \verb|Categorical_list_of_| & \verb|Categorical_list_of_| \\
       & \verb|  programming_languages| & \verb|  programming_languages| \\
     3 & \verb|C_programming_language| & \verb|C_programming_language| \\
     4 & \verb|Functional_programming| & \verb|Functional_programming| \\
     5 & \verb|Object-oriented_programming| & \verb|Object-oriented_programming| \\
     6 & \verb|Programming_paradigm| & \verb|Java_programming_language| \\
     7 & \verb|Java_programming_language| & \verb|Programming_paradigm| \\
     8 & \verb|Generic_programming| & \verb|Generic_programming| \\
     9 & \verb|Lisp_programming_language| & \verb|Lisp_programming_language| \\
    10 & \verb|Ada_programming_language| & \verb|Ada_programming_language| \\    
    \hline
   \end{tabular}}
   \label{table5}
  \end{center}
\end{table*}

\section{Conclusion} \label{conlusion}
\noindent We present a link structure ranking algorithm for trading network which is derived from analyzing the preferential attachment mechanism in the network. We show that the mechanism in trading network is opposite to the corresponding mechanism in WWW network induced from the HITS definition. The differences come from the fact that in trading network the links are the flows of the resources driven by the supply and demand principle, a fundamental principle behind the trading activities, and in WWW network the links are the hyperlinks that can be created without exchanging any resource. Because the preferential attachment is the underlying principle behind the HITS formulation, by utilizing the differences we are able to define a link structure ranking algorithm for trading networks. The distinct feature of our algorithm is the using of network structure in determining the ranking scores which is a popular method in the WWW network researches.

There are some possible applications of the proposed algorithm. The most obvious one is to use it as a metric to determine the degree of importance of agents involved in the trading activities. Different from the standard method of using aggregate transaction volumes, the proposed algorithm which makes use of the network structure will favour agents that are highly connected or link to (are linked by) other highly connected countries. Thus, the network structure which is an invaluable information in the graph theory but uncaptured in the standard method will become an essential factor in determining the degree of importance. The second possible application is to design a recommendation scheme; for example in online auction network where the number of users is enormous, the proposed algorithm will be helpful in focusing efforts to only the most important users that are relevant to the search queries. 

In the WWW network part, we show that the modified HITS which favours the preferential attachment in general has better convergence rates than the original HITS, thus it can be used to improve the HITS computations. This is an interesting subject on its own and has been studied in our other work. The readers can refer to (\citealp{56}) for detailed discussions.


\end{document}